\documentclass [final,5p,times,twocolumn]{elsarticle}

\usepackage{lineno}
\usepackage{appendix}
\modulolinenumbers[5]
\biboptions{sort&compress,square}
\journal{Nuclear Instruments and Methods in Physics Research A}
\usepackage{booktabs}
\biboptions{sort&compress,square}
\usepackage[english]{babel}
\usepackage{epstopdf}

\hypersetup{hidelinks} 

\begin{document}

\begin{frontmatter}

\title{Geant4 simulation of the ELIMED transport and dosimetry beam line for high-energy laser-driven ion beam multidisciplinary applications}



\author[a,b]{G. Milluzzo}
\author[a]{J. Pipek}
\author[a]{A.G. Amico}
\author[a]{G.A.P. Cirrone}
\author[a]{G. Cuttone}
\author[c]{G.Korn}
\author[a]{G. Larosa}
\author[a,b]{R. Leanza}
\author[c]{D. Margarone}
\author[a,b]{G. Petringa}
\author[a]{A. Russo}
\author[c,a]{F. Schillaci}
\author[c,a]{V. Scuderi}
\author[d,a]{F. Romano}

\address[a]{Istituto Nazionale di Fisica Nucleare, Laboratori Nazionali del Sud, Via Santa Sofia 62, Catania, Italy}
\address[b]{Universita di Catania, Dipartimento di Fisica e Astronomia, Via S. Sofia 64, Catania, Italy}
\address[c]{Institute of Physics ASCR, v.v.i (FZU), ELI-Beamlines project, 182 21 Prague, Czech Republic}
\address [d] {National Physical Laboratory, CMES - Medical Radiation Science, Teddington TW11 0LW, Middlesex, UK}





\begin{abstract}
The ELIMED (MEDical and multidisciplinary application at ELI-Beamlines) beam line  is being developed at INFN-LNS with the aim of transporting and selecting in energy proton and ion beams accelerated by laser-matter interaction at ELI-Beamlines in Prague.
It will be a section of the ELIMAIA (ELI Multidisciplinary Applications of laser-Ion Acceleration) beam line, dedicated to applications, including the medical one, of laser-accelerated ion beams \cite{Margarone2013, Cirrone2015a}
A Monte Carlo model has been developed to support the design of the beam line in terms of particle transport efficiency, to optimize the transport parameters at the irradiation point in air and, furthermore, to predict beam parameters in order to deliver dose distributions of clinical relevance.
The application has been developed using the Geant4 \cite{Agostinelli2003} Monte Carlo toolkit and has been designed in a modular way in order to easily switch on/off geometrical components according to different experimental setups and user’s requirements, as reported in \cite{Pipek2017}, describing the early-stage code and simulations. The application has been delivered to ELI-Beamlines and will be available for future ELIMAIA's users as \emph{ready-to-use} tool useful during experiment preparation and analysis. 
The final version of the developed application will be described in detail in this contribution, together with the final results, in terms of energy spectra and transmission efficiency along the in-vacuum beam line, obtained by performing end-to-end simulations. 

\end{abstract}
\end{frontmatter}

\section{Introduction}

The ELIMAIA beam line will be located in one of the future experimental halls at ELI Beamlines facility in Prague, i.e. the one dedicated to laser-driven ion acceleration as well as x-ray production and their possible multidisciplinary applications. \\
It will be composed of two sections: the first one is addressed to the Peta-Watt (PW) laser interaction with solid target, including two available interaction chambers, plasma mirrors and diagnostics, and the second one, called ELIMED, is a complete transport and dosimetry beam line dedicated to laser-driven ion beams handling, selecting and monitoring in view of possible multidisciplinary applications, such as the medical one. The latter has been in charge of INFN-LNS, which has signed a three-years contract with ELI Beamlines in 2014 to design, develop and realize the ELIMED beam line,including all the transport as well as dosimetry elements \cite{Cirrone2015,Romano2016}. 
The main goal is to provide stable and reproducible beams suitable for applications with a controlled acceptable energy spread (from 5$\%$ up to 20$\%$) and flux per shot and homogeneous  lateral and logitudinal dose distributions. 
The ELIMED beam line will be composed of two main transport magnetic elements: a set of five permanent magnet quadrupoles (PMQs), capable of focusing the ion species and the energy component to be transported \cite{Schillaci2015}, followed by an energy selector system (ESS) composed of 4 resistive dipoles and working as a single reference trajectory, capable to select the wanted energy, thanks also to a central slit with a variable width \cite{Schillaci2016}. 
The in-vacuum tranport section is followed by a final part in air, consisting of relative and absolute dosimeters, such as a secondary emission monitor (SEM), an in-transmission double gap ionization chamber (IC) and a Faraday cup (FC) \cite{Romano, Leanza}. Moreover, a specific Time Of Flight (TOF)-based diagnostics system, based on diamond as well as silicon carbide detectors, will be also used along the ELIMED beam line to provide online measurements of particle energy distribution and flux \cite{Milluzzo_1, Milluzzo_2, Scuderi}.\\
\noindent During the realization, a crucial role has been played by the Monte Carlo Geant4 simulation of the whole ELIMED beam line, which has supported the design of some key elements and the study and the optimization of beam transport parameters.
The Geant4 based application, reproducing all the transport and dosimetry elements, has been delivered to ELI-Beamlines and will be a tool available for future ELIMAIA's users interested in simulating their own experimental setup.\\
The final ELIMED application, developed at INFN-LNS, will be described in detail, together with some comparisons with results from reference transport code. Moreover, a dedicated study of the dependence of proton energy spectra and transmission efficiency downstream the in-vacuum section from the ESS central slit aperture will be also reported. 

\begin{figure*}[h!]
\centering
\includegraphics[width=350pt,height=140pt]{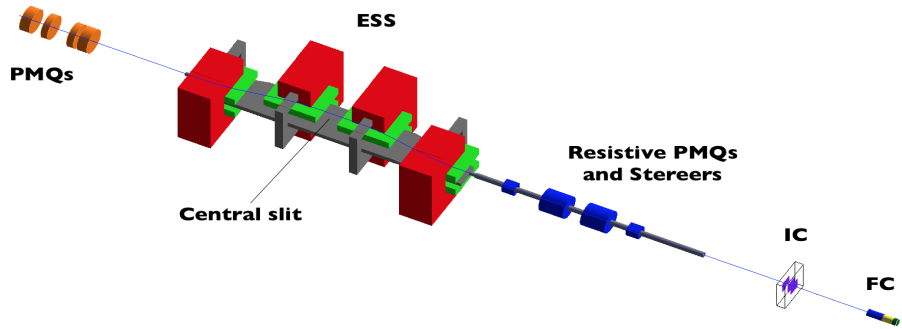}
\caption{ELIMED transport and dosimetry beam line simulated with Geant4 simulation toolkit, from the source down to the irradiation point. The quadrupole system (PMQs), the Energy Selector System (ESS) and the Central slit, the resistive PMQs and Stereers, the Ionization Chamber (IC) and the Faraday Cup (FC) are reported in figure.}
\label{sim}
\end{figure*} 
 
\section{The Geant4 ELIMED application}
The simulation software Geant4 (GEometry ANd Tracking) \cite{Allison,Allison2016}, based on the Monte Carlo method, is one of the most widely used Monte Carlo codes for particle interaction and transport in the matter. 
This toolkit is currently used in several fields, from High Energy
Physics to medical physics and space science, thanks to its advanced functionalities in the geometrical description and to a wide and well-tested set of physics models. Thanks to its robustness,
versatility and reliability of the implemented physical processes, Geant4 has been chosen as the most appropriate code for the ELIMED transport and dosimetry beam line simulation.\\
The application was developed following four main requirements:
\begin{itemize}
\item Realistically reproducing each single beam line element, in terms of geometry and magnetic features, maintaining a good flexibility of the code for user experimental setups
\item Implementing a realistic laser-driven ion source with specific energy and angular distribution. It is possible to use typical Particle In Cell simulation outputs as input of the simulation to obtain realistic results.
\item Giving the possibility to retrieve outputs at different positions along the beam line, including the final irradiation point
\item Providing a easy-to-use and friendly interface of the code for future users

\end{itemize}

In order to fulfill all these requirements, the ELIMED application has been conceived as complex and modular code, in which the different geometrical components, as well as their functionalities can be switched off/on according to the specific purpose of the simulation.
The scheme of the whole ELIMED beam line simulated with Geant4, indicating the main beam line components, is reported in figure \ref{sim}. 

\section{Transport beam line validation with reference optics analytic codes}
\label{benchmark}
\begin{figure}[h!]
\centering
\includegraphics[width=190pt,height=140pt]{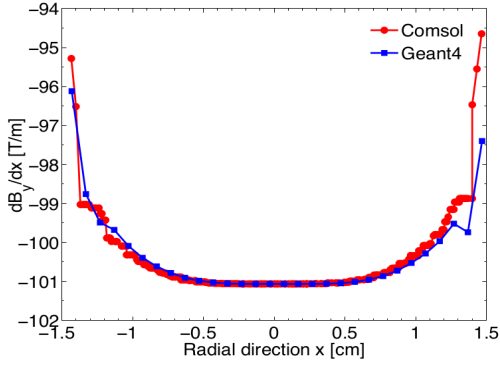}
\caption{Magnetic field gradient dBy/dx as a function of radial axis at the center of the 160 mm PMQ obtained with Geant4 (blue closed squares) and with COMSOL (red closed circles).}
\label{gradienti}
\end{figure}

\begin{figure}[h!]
\centering
\includegraphics[width=200pt,height=140pt]{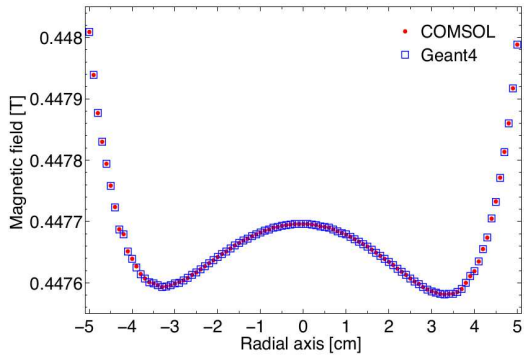} 
\caption{Magnetic field module along the radial axis at the center of the dipole: comparison between Geant4 (blue squares) and COMSOL (red circles).}
\label{BESS}
\end{figure}

\begin{figure}[h!]
\centering
\includegraphics[width=220pt,height=180pt]{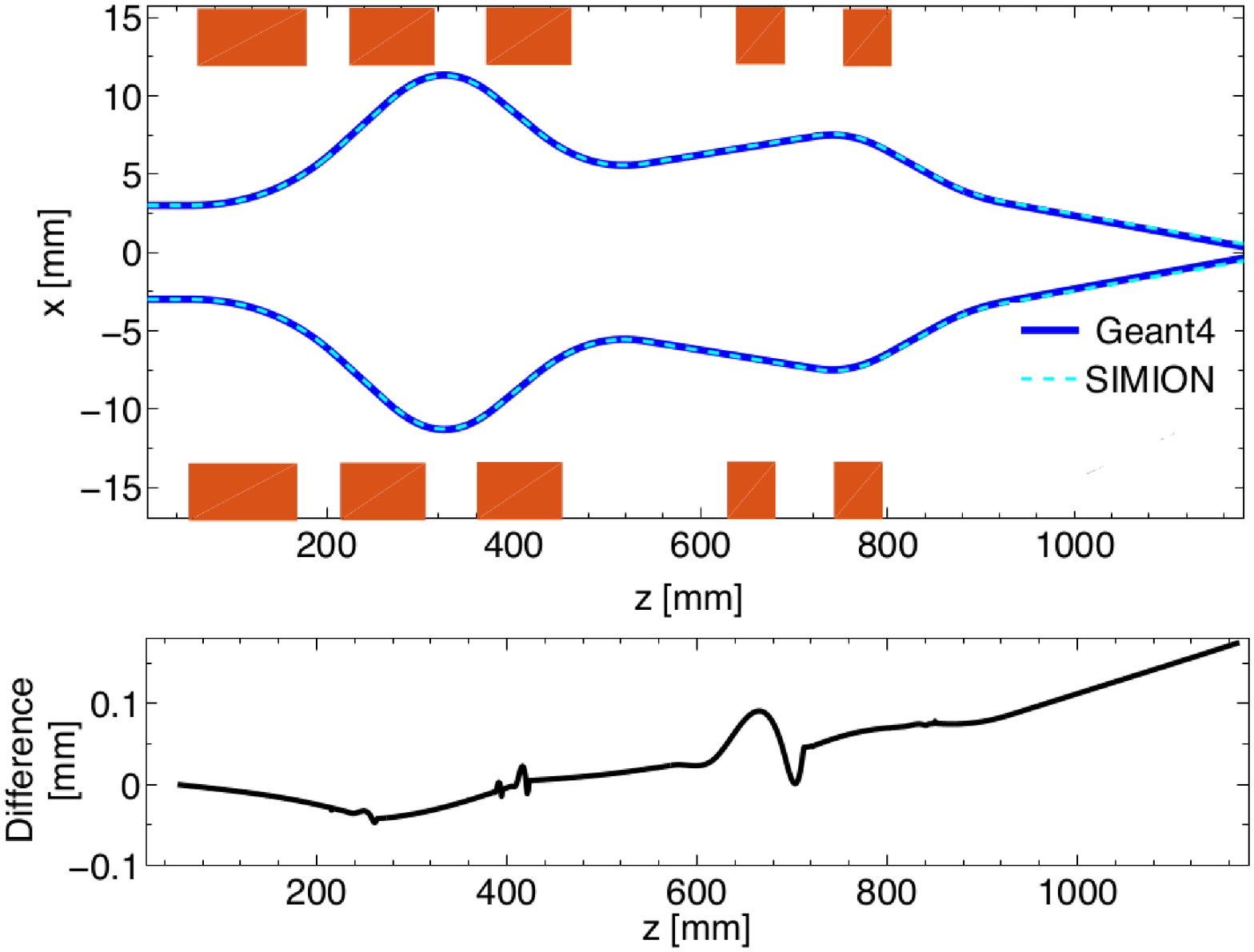}
\caption{Top: Comparison of the proton tracks simulated using SIMION (light blue dotted line) and Geant4 (blue line) within the focusing system. The projection of the tracks on the x-z plane for starting positions of x = $\pm$3 mm are shown. Bottom: Difference between the tracks obtained with Geant4 and SIMION along the PMQs system (black line).}
\label{pmqtrack}
\end{figure}

\begin{figure}[h!]
\centering
\includegraphics[width=220pt,height=140pt]{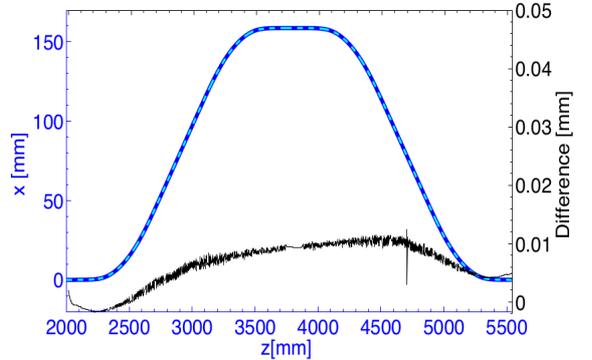}
\caption{Comparison of the 60-MeV proton trajectory within the ESS simulated by SIMION (light blue dotted line) and Geant4 (blue line) along the whole ESS. The difference between the tracks along the ESS in also reported (black line).}
\label{tracciaESS}
\end{figure}

The first transport element, is composed of five focusing high-gradient permanent quadrupoles (PMQs), whose configuration, i.e. relative distances, is determined according to the particle species and energy to be transported \cite{Schillaci2015}. The ELIMED Geant4 application includes two different implementations of the PMQs geometry, which can be selected  according to the purpose of the simulation: a simplified model in which the PMQs have been simulated as simple iron cylinders of different length, suitable when the purpose of the simulation is the transport optimization of primary beam along the whole beam line (figure \ref{sim}); a detailed model in which a stainless shielding layer is added around the PMQ bore and an internal structure of the magnet composed of alternating aluminium and neodymium part is included, more suitable to study secondary radiation production.
The PMQs system is followed by the Energy Selector System (ESS) placed in a in-vacuum chamber and composed by four resistive electromagnetic dipoles tunable from 0.17 up to 1.2 T, coupled with a central slit with variable aperture \cite{Schillaci2016}. The ESS geometric components have been reproduced in details in the Geant4 model. 
For both transport elements, to realistically simulate the PMQs focusing effect and the variable magnetic field of the ESS dipoles, 3D magnetic field maps can be imported in the application.
They have been obtained with the COMSOL \cite{Comsol} and OPERA \cite{opera} software for each quadrupole and for the whole ESS.
A dedicated class has been implemented, able to read the magnetic field components included in the imported map, interpolate the regular grid (typically 1 mm step) and simulate the magnetic field effect. 
The implementation of the magnetic fields in the ELIMED application has been checked by comparing the values obtained with Geant4 with the reference code COMSOL. 
Figure \ref{gradienti} shows the magnetic field gradient dBy/dx as a function of the radial direction (x) for the longer (160 mm length) PMQ obtained with Geant4 (blue squares) and with COMSOL (red circles). 
On the other hand, the magnetic field intensity at the center of the dipole along the radial direction again achieved for the two different codes is shown in figure \ref{BESS}. 
The good agreement obtained for both the magnetic field elements, demonstrates the correct implementation of the map within the Geant4-based application. \\
The particle tracking through the magnetic field obtained with Geant4 has been also benchmarked by comparing the proton trajectories along the five magnetic field and the ESS with the ones obtained, in the same configuration, with the reference transport code, SIMION \cite{simion}, used for optics studies. 
In particular, a specific configuration of the PMQs optimized to transport 60 MeV protons, obtained by means of the COMSOL software has been reproduced in the ELIMED application. 
Parallel 60 MeV proton beams with two different initial positions, i.e. ($\pm$3 mm, 0, 0), have been simulated and the particle position along the full trajectory has been registered step by step. Figure \ref{pmqtrack} (top) shows the trajectories obtained with Geant4 (blue line) and SIMION (light blue dotted line) in the x-z plane along the five PMQs, where x and z are respectively the radial and axial direction.
An optimal agreement between the tracks obtained with the two codes has been found, with a maximum discrepeancy of 150 $\mu$m as it is shown in figure \ref{pmqtrack} (bottom). \\
\noindent Similar comparisons have been carried out also for protons going through the ESS. In paricular a parallel 60 MeV proton beam has been simulated in the two codes, considering as a starting point the ESS entrance.
A magnetic field of 0.45 T has been set in both the simulations in order to transport 60 MeV protons according to the scaling law reported in \cite{Schillaci2016}. 
The comparison in figure \ref{tracciaESS} shows that the tracks are overlaped with a maximum deviation of about 10 $\mu$m at the slit position (figure \ref{tracciaESS}, i.e. at the symmetry center of the ESS, whose the width was fixed to 10 mm in the performed simulation. 
\\\
\section{In vacuum beam transport results}

\begin{figure}
\centering
\includegraphics[width=200pt,height=140pt]{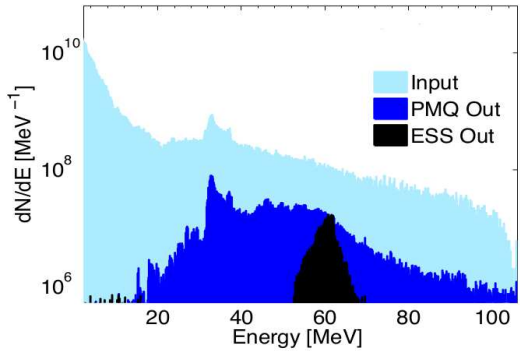}
\caption{Proton energy spectrum at the source (light blue), after the PMQs (blue) and dowstream of the ESS (black).}
\label{tutti}
\end{figure}

\begin{figure}[h!]
\centering
\includegraphics[width=200pt,height=150pt]{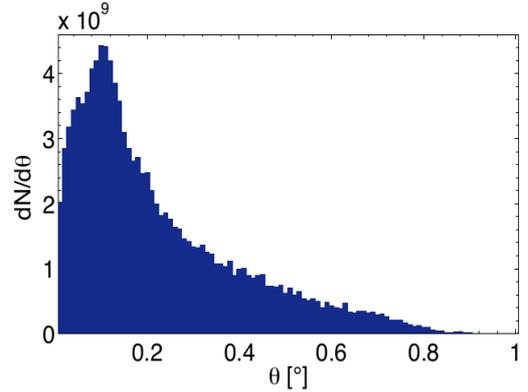}
\caption{Angular distribution ($\theta$) of protons in the energy range 60 $\pm$ 6 MeV after the quadrupole system.}
\label{ang}
\end{figure}

Once the reliability of the ELIMED application has been properly demonstrated as shown in section \ref{benchmark}, simulations from the source down to the in-air irradiation point have been performed in order to study the expected energy spectra and fluence per shot. 
Since experimental data are not yet available at the ELIMAIA beam line, Particle In Cell simulations have be used as input of the Geant4 simulations \cite{Psikal}. In
particular, two-dimensional PIC simulations have been performed, reproducing the interaction of 1 PW laser and a thin nanosphere target.
\noindent A typical exponential proton energy spectrum with a maximum energy of 100 MeV is reproduced, including the energy-angle dependence, reaching a maximum divergence of 60\,$^{\circ}$ at low energy ($<$ 20 MeV) and 20\,$^{\circ}$ at high-energy ($>$ 20 MeV). According to the PIC-2D simulation, the total number of protons per shot is 8.88 $\cdot 10^{11}$.
Nevertheless, $2.2 \cdot 10^7$ protons have been used as input of the performed Geant4 simulations, assuring a good statistics. The obtained results have been finally normalized to the number of protons per shot  given by PIC simulations, to obtain realistic predictions. 
Figure \ref{tutti} shows the normalized energy spectra retrieved on different planes along the beam line. In particular, the light blue one is related to the PIC2D laser-driven proton source energy spectrum, the blue one is the energy spectrum of the beam transmitted through the PMQs, i.e. just after the last quadrupole and the black one is the energy spectrum after the magnetic selection of the ESS, i.e. retrieved downstream the ESS. 
\noindent Both the PMQs and the ESS were set to optimize the transport of 60 MeV protons, using the same configuration reported for the code validation in section \ref{benchmark}.
As expected, a broad energy spectrum is still present downstream the PMQs, although they were set for 60 MeV proton transport. However, the focusing effect of the PMQs results in a lower divergence of protons with the \emph{selected} energy with respect to other energies, so that the injection in the ESS is optimized for the specific energy.
In agreement with the results reported in \cite{Schillaci2016} figure \ref{ang} shows the angular distribution of 60 $\pm$ 6 MeV protons downstream the PMQs, as they are injected in the ESS. A divergence of about 0.1$^{\circ}$ has been found, confirming the optimized injection in the ESS. 

\begin{figure}[h!]
\centering
\includegraphics[width=210pt,height=160pt]{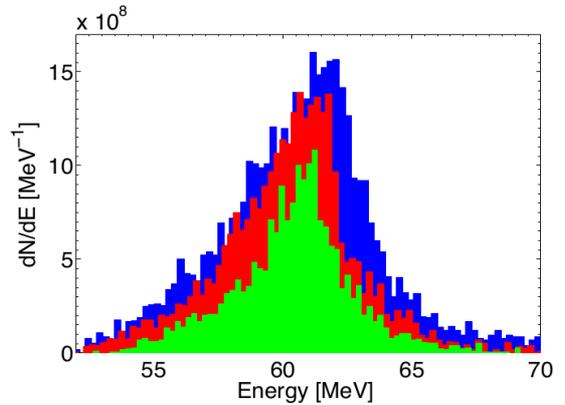}
\caption{Energy spectrum just for 60 MeV proton configuration downstream the ESS with a 20 mm (blue), 10 mm (red) and 5 mm (green)
slit aperture.}
\label{espectrum}
\end{figure}

As shown in figure \ref{tutti}, results obtained with the Geant4 application confirm that the ELIMED beam line will be able to deliver beams suitable for multidisciplicary applications, selecting in energy beams initially emitted with broad angular and energy distributions.
In particular, as already mentioned, the central slit width determines the energy resolution of particles reaching the end of the ESS. In order to investigate the energy spread dependence on the slit aperture,  simulations have been performed fixing the ESS magnetic field to 0.45 T, necessary to select 60 MeV protons, and varying the slit width (5 mm, 10 mm and 20 mm). 
The energy spectra registered downstream the ESS are shown in figure \ref{espectrum}. 
As expected, smaller slit apertures provide lower energy spread, ranging from 7 $\%$ up to 11$\%$ as reported in table \ref{spread60}. On the other hand, the slit width strongly affects the transmission efficiency, resulting higher for larger apertures than for narrower ones. The transmission efficiencies for 5, 10 and 20 mm slit width, evaluated as the ratio of the transmitted (output) and the emitted (input) protons within the energy range 60 $\pm$ 6 MeV are listed in table \ref{spread60}. 

\begin{table}[h!]
\centering
\begin{tabular}{ccc}
\toprule  
 Slit width [mm] & Energy spread & Transmission efficiency \\
\midrule
5  &  7 $\%$ & 5 $\%$  \\
10 &  8 $\%$ & 8 $\%$ \\
20 & 11 $\%$ & 10 $\%$ \\
\bottomrule
\end{tabular}
\caption{Energy spread and transmission efficiency for 60 MeV proton beams, considering  three different slit widths.}
\label{spread60}
\end{table}
In order to assure a limited energy spread and at the same time an acceptable number of particles transmitted per shot, still reasonable for multidisciplinary applications, a compromise has to be found also accoding to the specific experimental requirements.

\section{Conclusion}

\noindent The described Geant4-based application, reproducing in details geometrical as well as magnetic field features of the ELIMED transport and dosimetry elements, has been successfully validated comparing the field implementation and the tracks with the reference codes. Top-to-bottom simulations allowed to study the transported beam characteristics along the in-vacuum beam line, i.e. after the PMQs and at the ESS exit, confirming the influence of the slit width on the final energy spread and particle transmission efficiency. Further studies related to the final in-air section of the ELIMED beam line have been performed, including spatial dose distributions relevant for clinical applications and results will be presented in future works.

\section{Acknowledgments}
\noindent This work has been supported by the project ELI-Extreme Light Infrastructure-phase 2 ($CZ.02.1.01/0.0/0.0/15\_008/0000162$) from European Regional Development Fund, and partially by the projects  no. $CZ.1.07/2.3.00/20.008$7 and $CZ.1.07/2.3.00/20.0279$, which were co-financed by the European Social Fund and the state budget of the Czech Republic.

\bibliography{library}

\end{document}